\documentclass[a4paper,12pt]{article}
\usepackage{amsmath}
\usepackage{amssymb}
\usepackage{amsfonts}
\usepackage{fleqn}
\usepackage{graphicx}

\newcommand{\SuperField}[1]{\hat{#1}}

\def\<{\left\langle}
\def\>{\right\rangle}

\newcommand{\gsim}{\lower.7ex\hbox{$\;\stackrel{\textstyle>}{\sim}\;$}}
\newcommand{\lsim}{\lower.7ex\hbox{$\;\stackrel{\textstyle<}{\sim}\;$}}

\newcommand{\be}{\begin{equation}}
\newcommand{\ee}{\end{equation}}
\newcommand{\bea}{\begin{eqnarray}}
\newcommand{\eea}{\end{eqnarray}}

\begin{document}

\bibliographystyle{OurBibTeX}

\begin{titlepage}

\vspace*{-15mm}
\begin{flushright}
BA-06-14 \\
CERN-PH-TH/2006-071 \\
CAFPE/74/06\\ 
UG-FT/204/06 
\end{flushright}
\vspace*{5mm}

\begin{center}
{\bf \Large Supersymmetric Hybrid Inflation}
\\[2mm]
{\bf \Large with Non-Minimal K\"ahler potential }
\\[12mm]
M. Bastero-Gil\footnote{E-mail: \texttt{mbg@ugr.es}}$^{(a)}$,
S. F. King\footnote{E-mail: \texttt{sfk@hep.phys.soton.ac.uk}}$^{(b1,b2)}$,
Q. Shafi\footnote{E-mail: \texttt{shafi@bartol.udel.edu}}$^{(c)}$,
\\[5mm]

{\small\textit{$^{(a)}$
Departamento de Fisica Teorica y del Cosmos
and Centro Andaluz de Fisica de Particulas Elementales (CAFPE), %\\
Universidad de Granada, E-19071 Granada, Spain
}}
\\[3mm]

{\small\textit{$^{(b1)}$
TH Division, Physics Department,\\
CERN, 1211, Geneva 23\\
Switzerland
}}
\\[3mm]

{\small\textit{$^{(b2)}$
School of Physics and Astronomy,
University of Southampton,\\
Southampton, SO17 1BJ, U.K.
}}
\\[3mm]

{\small\textit{$^{(c)}$
Bartol Research Institute, University of Delaware, %\\
Newark, DE 19716, USA
}}
\\

\end{center}

\vspace*{2.0cm}

\begin{abstract}

\noindent
Minimal supersymmetric hybrid inflation based on a minimal
K{\"a}hler potential predicts a spectral index $n_s\gsim 0.98$.
On the other hand, WMAP three year data
prefers a central value $n_s \approx 0.95$.
We propose a class of supersymmetric
hybrid inflation models based on the same minimal superpotential
but with a non-minimal K{\"a}hler potential. Including
radiative corrections using the one-loop effective potential,
we show that the prediction for the spectral index is
sensitive to the small non-minimal corrections,
and can lead to a significantly red-tilted spectrum, in agreement with WMAP.
\end{abstract}

\end{titlepage}
\newpage
\setcounter{footnote}{0}

\section{Introduction}
Hybrid inflation models \cite{hybrid1,hybrid2} are examples of small
field inflation
models which predict a very small tensor fraction $r\ll 10^{-2}$.
Such models also typically predict an approximately
scale invariant spectral index.
For such models the WMAP three year central value for the spectral index is
about $n_s\approx 0.95$ \cite{Spergel:2006hy}, whereas the joint 
analysis of Ly-$\alpha$ forest power spectrum from the Sloan Digital
Sky Survey, with cosmic  microwave
background, galaxy clustering and supernovae yields $n_s =
0.965\pm 0.012$ \cite{Seljak}.
%combining the WMAP data with the SDSS galaxy survey \cite{Spergel:2006hy},
%\be
%n_s=0.948^{+0.015}_{-0.018}.
%\ee
%NOTE THAT WMAP HAVE MESSED UP THE ERRORS AND WILL BE ISSUING REVISED RESULTS.
%This red-tilted spectrum is significantly less than unity, in fact
%for such models $n_s=1$ is less preferred. 
%excluded at about 2$\sigma$
%\cite{Spergel:2006hy, Seljak}.
Consequently hybrid inflation models which predict the spectral index
to be too large are now less preferred \cite{Alabidi:2006qa}.

Amongst the models that are now less preferred by the WMAP three year
measurement of the spectral index are those based on
minimal supersymmetric hybrid inflation.
Minimal supersymmetric hybrid inflation may be defined by
the superpotential $W$,
\begin{eqnarray}\label{eq:W1}
\mathcal{W} &=& \kappa \SuperField{S}
(\SuperField{\phi} \SuperField{\bar \phi} - M^2) \,,
\end{eqnarray}
where $\SuperField{S}$ is a gauge singlet and
$\SuperField{\phi},\,\SuperField{\bar\phi}$ are a conjugate pair
of superfields
transforming as non-trivial representations of some gauge group $G$,
together with a minimal K{\"a}hler potential,
\be
\mathcal{K}_0=|S|^2 + |\phi|^2 + |\bar \phi|^2 \,,\label{minimal}
\ee
with $S$, $\phi$, $\bar \phi$ being the bosonic components of the
superfields. The gauge singlet $S$ is a natural candidate for the
inflaton in this model.
In the true supersymmetric minimum, $\phi$ and $\bar\phi$ have equal non-zero
vevs $\< \phi \> = \< \bar \phi \> = M$ whereas $\< S \> = 0 $ (or ${\cal O}
(m_{3/2})$ in broken supersymmetry).
During inflation, the theory is in a
false vacuum where $\< \phi \> = \< \bar \phi \> = 0$ and $\< S \> \not= 0$,
driving inflation. Inflation ends when the field value of the inflaton
$S$ falls
below some critical value which corresponds to a tachyonic instability for
$\< \phi \>$ and/or $\< \bar \phi \>$.
In this minimal model, the vevs $\< \phi \>$ and $\< \bar \phi \>$
break $G$ to some subgroup $H$. If $\phi,\bar{\phi}$ break
e.g.\ Pati-Salam or SO(10), topological defects are generated after inflation.
In order to avoid the monopole problem, one can extend superpotential to
so-called shifted \cite{shifted} or smooth inflation
\cite{smooth,Senoguz0512}, but here we shall restrict ourselves to the
minimal $W$ above.

The slow-roll parameters may be defined as
\begin{eqnarray}
\!\!\!\!\!\! \epsilon \!&=&\!
 \frac{m_\mathrm{P}^2}{2} \!\left(\frac{V'}{V}\right)^2  \,,\\
\!\!\!\!\!\!\eta \!  &= &\!   m_\mathrm{P}^2\!\left(\frac{V''}{V}\right) \,,\\
\!\!\!\!\!\!\xi^2 \!  &=&\!   m_\mathrm{P}^4 \!\left(\frac{V'\,
 V'''}{V^2}\right) \,,
\end{eqnarray}
where $m_\mathrm{P}=2.4\times 10^{18}$ GeV is the reduced Planck
mass.
Assuming that the slow-roll approximation
is justified (i.e.\ $\epsilon \ll 1$, $\eta \ll 1$), the spectral index
$n_\mathrm{s}$, the tensor-to-scalar ratio
$r=A_\mathrm{t}/A_\mathrm{s}$ and the running of the spectral index
$\mathrm{d} n_\mathrm{s}/\mathrm{d} \ln k$ are given by
\begin{eqnarray}
n_\mathrm{s} \!&\simeq&\! 1 - 6 \epsilon + 2 \eta \,,\\
r \!&\simeq&\! 16 \epsilon \,,\\
\frac{\mathrm{d} n_\mathrm{s}}{\mathrm{d} \ln k} \!\!&\simeq&\!\!
16 \epsilon \eta - 24 \epsilon^2 - 2 \xi^2 \,. 
\end{eqnarray}

The theory defined above in Eqs. (\ref{eq:W1}) and  (\ref{minimal})
defines the minimal supersymmetric hybrid inflation model.
As we shall see in the next section, it leads to a prediction
for the spectral index which is rather close to unity,
$n_s\gsim 0.98$, which is larger than the central value preferred by
WMAP three year data. On the other hand there is no symmetry that
protects the minimal form of the K{\"a}hler potential.
In this paper we study supersymmetric hybrid inflation with
non-minimal K{\"a}hler potential, including radiative corrections
using the one-loop effective potential, and show that the prediction
of the spectral index is sensitive to such non-minimal effects,
which can lead to a significantly red-tilted spectrum. This is done in
Section 3. The summary is presented in the last section, where we also
briefly comment on reheating after inflation.

\section{Minimal K{\"a}hler Potential}
In supersymmetric theories based on supergravity, there is a well
known problem that $\eta \approx 1$ due to the supergravity corrections,
thereby violating one of the slow roll conditions, and leading to
the so-called $\eta$ problem \cite{eta}.
It is an interesting fact that
the supergravity potential based on the minimal supersymmetric hybrid inflation
theory defined in Eqs.(\ref{eq:W1}), (\ref{minimal}) provides a solution to the $\eta$ problem
since the mass squared of the inflaton when calculated from the
supergravity potential cancels at the tree level.

In general the supergravity potential, including just the F-terms, is:
\begin{eqnarray}
V_\mathrm{F} = e^{\mathcal{K}/m^2_\mathrm{P}} \left[
K_{ij}^{-1} D_{z_i} \mathcal{W} D_{z^*_j} \mathcal{W}^*
- 3 m_\mathrm{P}^{-2} |\mathcal{W}|^2
\right] ,
\end{eqnarray}
with $z_i$ being the bosonic components of the superfields
$\SuperField{z}_i \in \{ \SuperField{\phi},\SuperField{S},\dots\}$
and where we have defined
\begin{eqnarray}
D_{z_i} \mathcal{W} := \frac{\partial  \mathcal{W}}{\partial z_i} +
 m_\mathrm{P}^{-2} \frac{\partial  \mathcal{K}}{\partial z_i} \mathcal{W}
 \: , \; K_{ij} := \frac{\partial^2 \mathcal{K}}{\partial z_i \partial
 z^*_j} \,,
\end{eqnarray}
and $D_{z^*_j} \mathcal{W}^* := (D_{z_j} \mathcal{W})^*$.
For the superpotential in Eq. (\ref{eq:W1}) and the
minimal K{\"a}hler potential $\mathcal{K}_0$
in Eq. (\ref{minimal}), the supergravity potential leads to:
%\begin{eqnarray}
%V_\mathrm{F} = e^{\mathcal{K}_0/m^2_\mathrm{P}} \left[
%D_{z_i} \mathcal{W} D_{z^*_i} \mathcal{W}^*
%- 3 m_\mathrm{P}^{-2} |\mathcal{W}|^2
%\right] ,
%\end{eqnarray}
%where
%\begin{eqnarray}
%D_{z_i} \mathcal{W} := \frac{\partial  \mathcal{W}}{\partial z_i} +
% m_\mathrm{P}^{-2} z^*_i\mathcal{W}
%\end{eqnarray}
%This leads to the tree-level potential,
\begin{eqnarray}
V_0^{\mathrm{min}} &\simeq& 2 \kappa^2 |S|^2 |\phi|^2 + \kappa^2 \left( |\phi |^2-M^2
\right)^2 \left(1 + 2 \frac{|\phi |^2}{m_\mathrm{P}^2} +  \frac{|S |^4}{2
  m_\mathrm{P}^4} + \frac{|\phi |^4}{
  m_\mathrm{P}^4}  \right) + \cdots  \label{V0min}
\,,
\end{eqnarray}
and we see that to leading order in the supergravity expansion
the mass squared term for the inflaton field $S$ has canceled.
This is fortunate since, if present, we would expect such a
supergravity induced mass squared to have the same form as the $\phi$
mass squared\footnote{The $\phi$ and $\bar{\phi}$ fields also receive
 field dependent masses during inflation given by the $S$ field.
 Other squark, slepton and Higgs fields are not included explicitly but
 are necessarily present in any realistic supersymmetric model.
 Such fields would be
 expected to get masses of order the Hubble constant during inflation,
 effectively lifting all flat directions involving these fields,
 which therefore play no role in inflation.}, 
 namely
$\kappa^2 M^4/m_\mathrm{P}^2=V_0/m_\mathrm{P}^2$ which is
of order the Hubble constant squared $H^2=V_0/3m_\mathrm{P}^2$.
The fact that the inflaton acquires a mass of order the Hubble constant
is a generic feature of supergravity, and gives rise to
$\eta \approx 1$, violating the slow roll condition
and leading to the so-called $\eta$ problem.
However, as already noted, in minimal supersymmetric hybrid inflation
above the mass squared term for the inflaton $S$ cancels and there is
no $\eta$ problem.

Since the tree-level mass squared for the inflaton $S$ cancels,
in minimal supersymmetric hybrid inflation the curvature
of the potential is given by the 1-loop effective potential,
\be
V_1^{\mathrm{min}}= V_0^{\mathrm{min}} + \Delta
\mathcal{V}_{\mathrm{1loop}} \,.
\ee
with the radiative correction  given by
\begin{eqnarray}
\Delta \mathcal{V}_{\mathrm{1loop}} = \frac{1}{64 \pi^2} \,\,\mbox{Str}\,
\left[ \mathcal{M}^4 (\phi ) \left( \ln\frac{\mathcal{M}^2
(\phi ) }{Q^2}-\frac{3}{2}\right) \right] \,,
\end{eqnarray}
where $\mathcal{M}^2 (\phi )$ is the
field-dependent mass-squared matrix of the contributing particles,
i.e., $\phi$ and $\bar \phi$, and $Q$ the renormalization
scale\footnote{None of the derivatives of $\Delta
  \mathcal{V}_{\mathrm{1loop}}$ depend on the renormalization scale
  $Q$, and therefore it would have no effect on the inflationary
  predictions. We can always choose for example this scale such that
  it minimizes  the value of $\mathcal{V}_{\mathrm{1loop}}$ along the
  inflationary trajectory. }.
The leading contributions of the 1-loop effective potential
can be expressed analytically as:
\begin{eqnarray}
\Delta \mathcal{V}_{\mathrm{1loop}}&\simeq& \frac{(\kappa M)^4}{8
  \pi^2} {\cal N} F[x] \,,\label{V1loop}\\
F[x] &=& \frac{1}{4}\left( (x^4+1) \ln \frac{(x^4-1)}{x^4}+2 x^2
  \ln \frac{x^2+1}{x^2-1} + 2 \ln \frac{\kappa^2 M^2 x^2}{Q^2}-3 \right) \,,
\end{eqnarray}
${\cal N}$ being the
dimensionality of the representation of the fields $\phi$, $\bar \phi$,
and where we have defined $x= |S|/M$.

During the inflationary epoch where $|S|>|S^{c}|= M$ the
waterfall field $\phi$ is held at zero due to its having a large
positive mass squared,
then when $S$ reaches $S^{c}$ the waterfall field $\phi$ rolls
out towards its global minimum, effectively ending inflation in the usual
way in hybrid inflation. Writing the potential in terms of the real
field $S_R= \sqrt{2} |S|$, and setting $|\phi|=0$, effectively during
inflation we are left with the potential\footnote{There is also a soft
mass term for the inflaton in the potential, $m_{3/2}^2 |S|^2$,
typically of the order of $m_{3/2} \simeq O(1 \, \mathrm{TeV})$, but
this term is only relevant for values of the coupling $\kappa < 10^{-5}$,
which we do not consider in this letter.}:
\be
V= V^{\mathrm{min}}_1 (\phi=0) \simeq \kappa^2 M^4 \left( 1 + \frac{S_R^4}{8
  m_\mathrm{P}^4} + \cdots \right)
+\Delta \mathcal{V}_{\mathrm{1loop}} \,.
\ee
As far as we have inflation for field values well below the Planck
scale, and $\kappa \gsim 10^{-3}$,
we can neglect the  quartic correction induced by the sugra
correction\footnote{Although this will become relevant for values of
  the coupling $\kappa \gsim 0.05$, see later. For values of the
  coupling $\kappa \ll 10^{-3}$, the potential is extremely flat
  and observable inflation takes place quite near the critical
  value. In the limit
  $|S|/M \rightarrow 1$, the sugra term dominates again over the
  radiative contribution \cite{Senoguz0412,Postma}.}.
The slow-roll parameters are then given:
\begin{eqnarray}
\epsilon &\simeq&  \frac{\kappa^2}{(4 \pi)^2} \left( \frac{\kappa
  m_\mathrm{P}}{4 \pi
  M}\right)^2 {\cal N}^2 F^\prime[x]^2 \,,\label{epsh} \\
\eta &\simeq& -\delta \simeq \left( \frac{\kappa m_\mathrm{P}}{4 \pi
  M}\right)^2 {\cal N} F^{\prime \prime}[x] \label{etah} \,,
\end{eqnarray}
where we have denoted by $\delta$ the
contribution to $\eta$ from the effective potential. The functions
$F^\prime[x]$ and $F^{\prime \prime}[x]$ are the first and
second derivative of $F[x]$ respectively, which
for $x > 1$ behave like $F^\prime[x] \simeq 1/x$, $F^{\prime
  \prime}[x] \simeq -1/x^2$. Therefore, in
that regime we have approximately:
\bea
\eta &\simeq&  - \left( \frac{\kappa m_\mathrm{P}}{4 \pi
  M}\right)^2 \frac{{\cal N}}{x^2}  \label{delta} \,,\\
\epsilon &\simeq& \frac{\kappa^2}{(4 \pi)^2} \delta \ll |\eta| \,,
\eea
The spectral index is then :
\be
n_s\simeq 1 - 2\delta  \,. \label{spectral}
\ee
The amplitude of the primordial spectrum is given by:
\begin{equation}
P_{\cal R}^{1/2} \simeq \frac{V}{V^\prime} \left(\frac{H}{2\pi
m_{\mathrm{P}}^2}\right)
\simeq \frac{1}{\sqrt{2 \varepsilon}} \left(\frac{H}{2\pi
m_{\mathrm{P}}}\right)
\simeq \sqrt{\frac{2}{3}}
  \left(\frac{4 \pi}{\kappa} \right) \left( \frac{M}{m_\mathrm{P}}\right)^3
  \frac{x_e}{{\cal N}} \,,
\label{spectrum}
\end{equation}
evaluated for the field value $x_e=S_{Re}/(\sqrt{2}M)$ at
$N_e$ e-folds before the end of inflation,
\be
N_e = \int_{S_{Re}}^{{S_R^c}} H dt \simeq
\int_{S_R^c}^{S_{Re}}  \frac{3 H^2}{V^\prime}d S_R
\simeq \left(\frac{4 \pi
  M}{\kappa m_\mathrm{P}}\right)^2 \int_{1}^{x_e}
\frac{dx}{F^\prime[x] {\cal N}} \,,
\label{Ne}
\ee
which again when $x_e > 1$ can be approximated by:
\begin{equation}
S_{Re} \simeq \sqrt{N_e {\cal N}}\frac{\kappa}{2 \pi} m_\mathrm{P} \,.
\label{phinue}
\end{equation}
Finally, using Eq. (\ref{phinue}) into Eq. (\ref{spectrum}), we get
the predicted amplitude of the primordial spectrum at $N_e$:
\be
P_{\cal R}^{1/2} \simeq 2 \sqrt{\frac{N_e}{3}} \left(
\frac{M}{m_\mathrm{P}}\right)^2 \,, \label{spectrum2}
\ee
The WMAP normalization is $P_{\cal R}^{1/2}= 4.86 \times 10^{-5}$,
taken at the comoving scale $k_0=0.002$ Mpc$^{-1}$. During inflation,
this scale exits the horizon at approximately \cite{leach}:
\be
N_e \simeq 53 + \frac{1}{3}\ln (T_R/10^9
\mathrm{GeV})+ \frac{2}{3} \ln (\sqrt{\kappa} M/10^{15} \mathrm{GeV}) \,,
\ee
where $T_R$ is the reheating temperature, which for $\kappa \geq
10^{-3}$ is expected to be of the order of $O(10^9\, \mathrm{GeV})$
\cite{Senoguz0412}.
This sets\footnote{For the numerical results
shown in the figures, varying $\kappa$ and taking $T_R \simeq 10^9$
GeV, we have checked that we always stay in the range $N_e \simeq
50-53$.}  $N_e \approx 50$,
and from Eq. (\ref{spectrum2}) this fixes the inflationary
scale $M \simeq 6 \times 10^{15}$ GeV.
Thus  for the spectral index, using Eqs. (\ref{phinue}) and (\ref{delta}),
we have the approximated result \cite{hybrid2}:
\be
n_s \simeq 1 - \frac{1}{N_e} \simeq 0.98 \label{spectralNe}\,,
\ee
for $N_e \approx 50$. The tensor to scalar ratio is negligible, with
$r \lsim 10^{-4}$, and also there is no running in the spectral index,
with $d n_s/d \ln k \lsim 10^{-3}$ \cite{Senoguz0309}.

\begin{figure}[th]
\hfil\scalebox{0.6} {\includegraphics{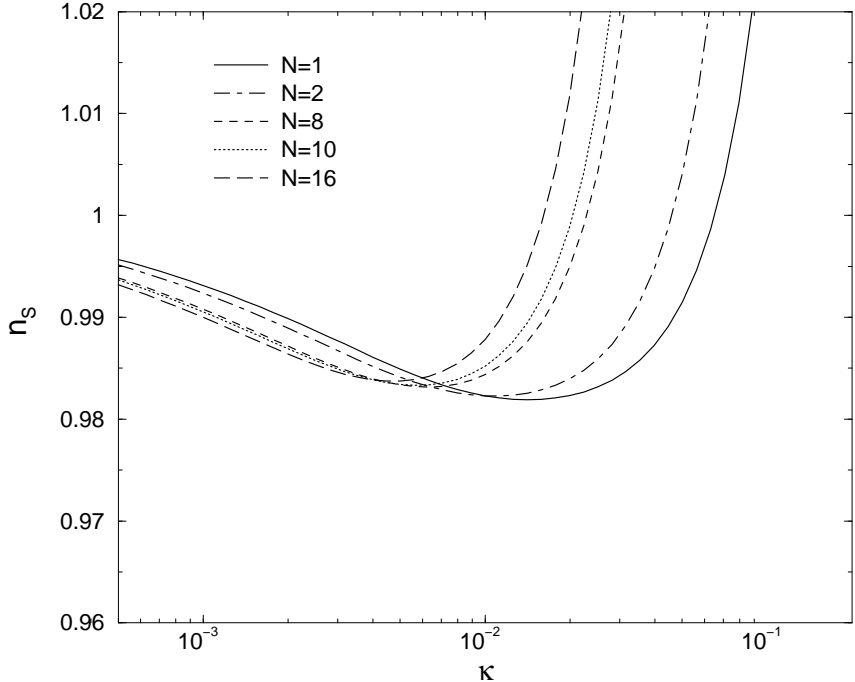}}\hfil
\caption{{\footnotesize Predicted value of the spectral index $n_s$
  depending on the   value of the coupling $\kappa$, for different
  values of ${\cal N}=1,2,8,10,16$. We have set $N_e = 50$. Varying
  the no. of e-folds up    to say $N_e=60$ would increase the
  predicted value of $n_s$ by at   most $0.005$.}}
\label{plot1}
\end{figure}
We can obtain in the same way the value $n_s$ in general
without making use of any approximations: (i) having chosen the no. of
e-folds at which  the primordial spectrum is normalised,  from
Eq. (\ref{Ne}) one obtains the corresponding value of the field
$S_{Re}$; (ii) knowing the value of the field, the WMAP  normalization
on the spectrum Eq. (\ref{spectrum}) fixed the scale of inflation $M$;
(iii) finally, Eqs. (\ref{etah}) and (\ref{spectral}) gives the
predicted value of the spectral index as a function of $\kappa$.
The predicted value of the spectral index is plotted in
Fig. (\ref{plot1}), showing the deviations from the approximated value
Eq. (\ref{spectralNe}) for small and large values of $\kappa$.
For small values of the coupling $\kappa$,
the approximation $x_e > 1$ does not
hold. Diminishing the coupling what we have is a flatter potential,
with a smaller curvature, so that the last say 50 e-folds of inflation
happens to be quite close to the critical value, with $x_e \approx 1$,
giving rise to a practically scale invariant spectrum. On the other
hand, for larger values of the coupling $\kappa$ although the
approximation $x_e > 1$ holds, one can see from Eq. (\ref{phinue})
that the value gets larger and closer to the Planck scale, so that the
quartic term for the inflaton induced by the  sugra corrections cannot
be neglected any longer. This tends to give a positive curvature
contribution, making the spectrum to turn from red tilted ($n_s <1$)
to blue tilted ($n_s > 1$).

The result in Eq. (\ref{spectralNe}) can be viewed as the lower bound
on the predicted spectral index, with  $n_s \gsim 0.98$,
to be compared to the central WMAP three year central value of $n_s
\approx 0.95$. This motivates supersymmetric
hybrid inflation with a non-minimal K{\"a}hler potential, where the
spectral index can be lowered.

\section{Non-Minimal K{\"a}hler Potential}

We now turn to the non-minimal modification of supersymmetric
hybrid inflation.
We continue to assume the same minimal superpotential as in Eq.(\ref{eq:W1}).
However we now consider a non-minimal K{\"a}hler potential,
\cite{Senoguz:2004ke,Antusch:2004hd,Seto},
\begin{eqnarray}\
\!\!  \mathcal{K}\!\!\!\! &=& \!\!\!\!
|S|^2 + |\phi|^2 + |\bar \phi|^2
+\kappa_S \frac{|S|^4}{4m_\mathrm{P}^2}
+\kappa_{S\phi} \frac{|S|^2 |\phi|^2}{m^2_\mathrm{P}}
+ \kappa_{S\bar \phi} \frac{|S|^2  |\bar
    \phi|^2}{m^2_\mathrm{P}}
+ \kappa_{SS} \frac{|S|^6}{ 6 m_\mathrm{P}^4}+ \cdots  \,.
\label{Knonmin}
\end{eqnarray}
Working along the D-flat direction $|\phi| = |\bar \phi|$,  and
keeping the relevant terms for
inflation up to $O((|S|/m_\mathrm{P})^{4})$, we get the potential:

\begin{eqnarray}
V_1^{\mathrm{non-min}} \!\!\!&\simeq& \!\!\! V^{\mathrm{non-min}}_0 +
\Delta\mathcal{V}_{\mathrm{1loop}}   \,, \nonumber \\
V^{\mathrm{non-min}}_0\!\!\!  &\simeq& \!\!\!
2 \kappa^2 |S|^2 |\phi|^2 \nonumber \\
\!\!\!\!\!\! &+&\!\!\!\kappa^2 \left(|\phi|^2-M^2 \right)^2
\left(1 - \kappa_S \frac{|S|^2}{m_\mathrm{P}^2}
+\kappa_\phi \frac{|\phi|^2}{m_\mathrm{P}^2}
+\gamma_S \frac{|S|^4}{2 m_\mathrm{P}^4} \right)+\cdots  \,,
\end{eqnarray}
where $\Delta\mathcal{V}_{\mathrm{1loop}}$ is
given in Eq. (\ref{V1loop}), and we have defined:
\bea
\kappa_\phi &=& (1-\kappa_{S\phi}- \kappa_{S\bar \phi}) \,,\\
\gamma_S&=& (1- \frac{7  \kappa_S}{2} + 2 \kappa_S^2 - 3 \kappa_{SS})
\,. %\\
%\gamma_\phi &=& \,.
\eea
The non-minimal
K{\"a}hler only introduces a small correction to the $\phi$ squared
mass, so that still for values of the inflaton field $|S| > |S|^c$
this is positive and we can set $\phi=0$ during inflation, with:
\bea
\!\!\! V \!\! = \!\! V_1^{\mathrm{non-min}} (\phi=0)\!\!\! &\simeq& \!\!\!  \kappa^2 M^4 \left(1 - \kappa_S
\frac{S_R^2}{2 m_\mathrm{P}^2}
+ \gamma_S \frac{S_R^4}{8 m_\mathrm{P}^4} +\cdots \right)
+\Delta \mathcal{V}_{\mathrm{1loop}} \,. \label{nonminpot}
\end{eqnarray}
Although it seems that we are introducing an infinite number
of arbitrary parameters in the expansion of the K\"ahler potential,
Eq. (\ref{Knonmin}),
we remark that in the regime where the inflaton field value is well
below the Planck mass, the non-minimal K\"ahler contributions to the
quartic and
higher terms for the inflaton have no effect on the inflationary
dynamics and therefore
only one parameter, $\kappa_S$, will be relevant for the inflationary
predictions that  follow. Note that $\kappa_S >0$ will be required so
that the prediction for $n_s$ is in agreement with WMAP.

%Given the potential, we can follow the standard procedure sketched in
%the previous section, getting first the value of the field at $N_e$
%e-folds and then imposing the WMAP normalisation on the spectrum in
%order to get the scale $M$, for each pair of values $\kappa_S$,
%$\kappa$.
The non-minimal K{\"a}hler induces now a negative
correction to both the first and the second derivative of the
potential in the inflaton direction:
\bea
V^\prime &\simeq& \frac{\kappa^2 M^4}{m_\mathrm{P}}
\left( -\kappa_S \frac{S_R}{m_\mathrm{P}}
+ \gamma_S \frac{S_R^3}{2 m_\mathrm{P}^3} +
\frac{\kappa^2 m_\mathrm{P}}{8 \sqrt{2} \pi^2 M } {\cal N}
F^\prime[x] \right) \,,
\label{Vprime}\\
V^{\prime \prime} &\simeq& \frac{\kappa^2 M^4}{m_\mathrm{P}^2} \left(
-\kappa_S +
3 \gamma_S \frac{S_R^2}{2 m_\mathrm{P}^2} +
\frac{\kappa^2 m_\mathrm{P}^2}{16 \pi^2 M^2} {\cal N} F^{\prime
  \prime}[x] \right) \,.
\eea
This correction gives rise to a local minimum and maximum in the
potential located at
\bea
\frac{S_R^{min}}{m_\mathrm{P}} &\simeq& \sqrt{\frac{2\kappa_S}{\gamma_S}}
  \,, \label{SRmax} \\
\frac{S_R^{max}}{m_\mathrm{P}} &\simeq& \sqrt{\frac{2 {\cal
      N}}{\kappa_S}}\left(\frac{\kappa}{4 \pi}\right) \,, \label{SRmin}
\eea
which for example for $\kappa_S \simeq \kappa \simeq 0.01$ gives
$S_R^{min}/m_\mathrm{P} \simeq 0.14$, and $S_R^{max}/m_\mathrm{P}
\simeq 0.01$.  After that, for $S_R < S_R^{max}$ we have the standard
flat potential with $V^\prime >0$,  suitable for  hybrid inflation,
with the field rolling  towards the critical value.
We have demanded then that we can get at least 60-50 e-folds of
inflation once the field is in that region of the
potential with $V^\prime >0$,  i.e., that $S_{Re} \leq S_R^{max}$.
We do not address the question of how the field reaches $S_{Re}$ in
this letter, that is, the problem of the initial conditions for
inflation. Although this problem is also present to some extent in the
minimal case, it can be more severe in the non-minimal scenario due to
the presence of the local minimum near Planck values. Starting the
evolution for the homogeneous inflaton field near or beyond Planck, it
may happen that the field gets stuck in this local minimum, and the
system may inflate there, but it is not clear how to end inflation. On
the other hand depending on the initial field values, the field may
overcome the minimum, reach the flat part of the potential, and hybrid
inflation may start. We can always find such initial values at least
for the inflaton field, but then the question would be how fine-tuned
they are. Nevertheless, when studying the evolution of the
system prior to inflation, fields such as $\phi$, $\bar \phi$,
should be taken into account. One should also check that these fields
indeed go early enough to their respective local minimum. This is an important issue, but beyond the
scope of this letter. Here we just concentrate on the inflationary
predictions derived from the potential Eq. (\ref{nonminpot}), assuming
that we have suitable initial conditions for hybrid inflation to take
place.

\begin{figure}[t]
\hfil\scalebox{0.6} {\includegraphics{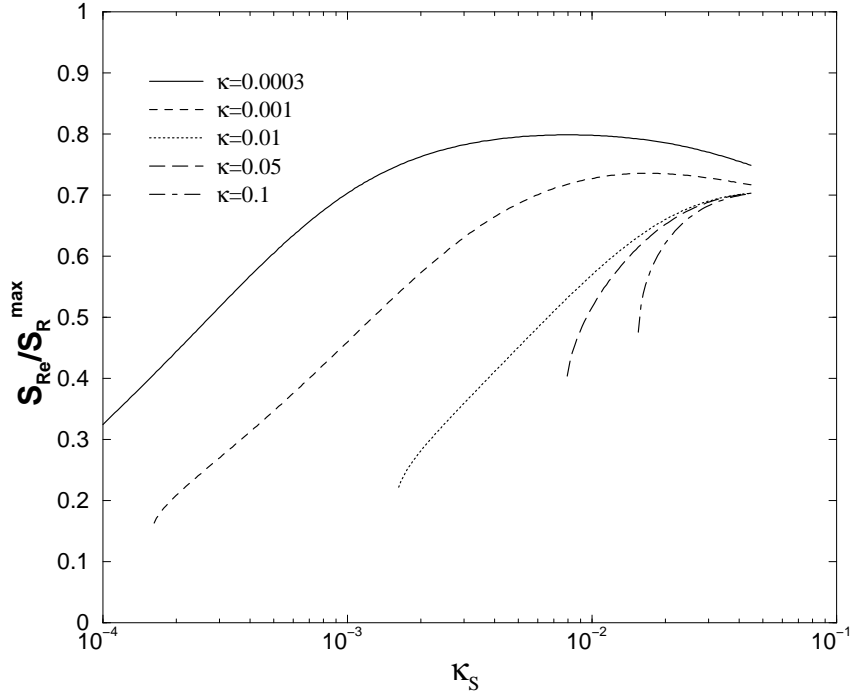}}\hfil
\caption{{\footnotesize Ratio $S_{Re}/S_R^{max}$ versus $\kappa_S$, for
  different values of $\kappa$. }}
\label{plotratio}
\end{figure}

From Eq. (\ref{SRmax}), the condition of having enough inflation,
$S_{Re} \leq S_R^{max}$, might be expressed as an upper bound on
the possible value of $\kappa_S$, with
\be
\kappa_S \lsim 2 {\cal N}\left(\frac{\kappa}{4 \pi}\right)^2 \left(
\frac{m_\mathrm{P}}{S_{Re}} \right)^2 \label{bound}\,.
\ee
However, the value of the field at $N_e$ e-folds given by
Eq. (\ref{Ne}), $S_{Re}$,
itself depends on the value of $\kappa_S$ through $V^\prime$. The
contribution from $\kappa_S$ tends to decrease $V^\prime$ and makes
the potential flatter, so that the corresponding value of $S_{Re}$
decreases and it will stay below $S_R^{max}$. This is shown in Fig.
(\ref{plotratio}), where we have plotted the ratio $S_{Re}/S_R^{max}$
depending on $\kappa_S$, for different values of $\kappa$. As
$\kappa_S$ increases the ratio also increases but remains below one. On the
other hand, the minimum/maximum in the potential disappears whenever $
\kappa_S \lsim \sqrt{\gamma_S {\cal N}} \kappa/(2 \pi)$,
this being the lower value of $\kappa_S$ shown for each curve in
Fig. (\ref{plotratio}).

In addition, we have now in Eq. (\ref{nonminpot}) a mass term for the
inflaton field proportional to $\kappa_S$, and therefore this
parameter has to be small enough in order to satisfy the slow-roll
conditions.  The slow-roll parameters are given by:
\begin{eqnarray}
\epsilon &\simeq& \left( -\kappa_S \frac{S_R}{m_\mathrm{P}} +
  \left(\frac{\kappa^2}{4 \pi}\right)^2 \left(
  \frac{m_\mathrm{P}}{M}\right) {\cal N}F^\prime[x] \right)^2  \,,\\
\eta &\simeq& -\kappa_S - \delta \,,
\end{eqnarray}
where we have assumed $S_R \ll m_\mathrm{P}$ so that we can neglect
the quartic term in the analytical expression\footnote{The quartic
  term for the inflaton is taken into account in all the numerical
  calculations, and therefore in the results presented in the plots.},
$\delta$ is the contribution
from the 1-loop effective potential, Eq. (\ref{etah}), and again
$\epsilon \ll |\eta|$.
Therefore, for slow-roll inflation, $|\eta| <1$, we only
require $\kappa_S <1$.

\begin{figure}[t]
\hfil\scalebox{0.6} {\includegraphics{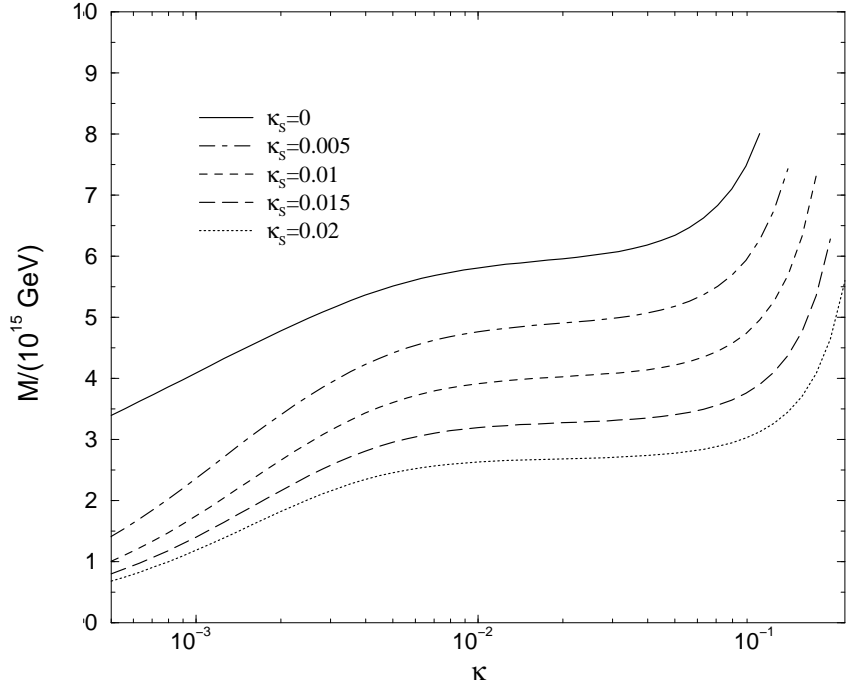}}\hfil
\caption{{\footnotesize Value of $M$ depending on $\kappa$, for
  different values of $\kappa_S$; from top to bottom
  $\kappa_S=0,\,0.005,\,0.01,\,  0.015,\,0.02$. (${\cal N}=1$) } }
\label{plotM}
\end{figure}

\begin{figure}[th]
\hfil\scalebox{0.6} {\includegraphics{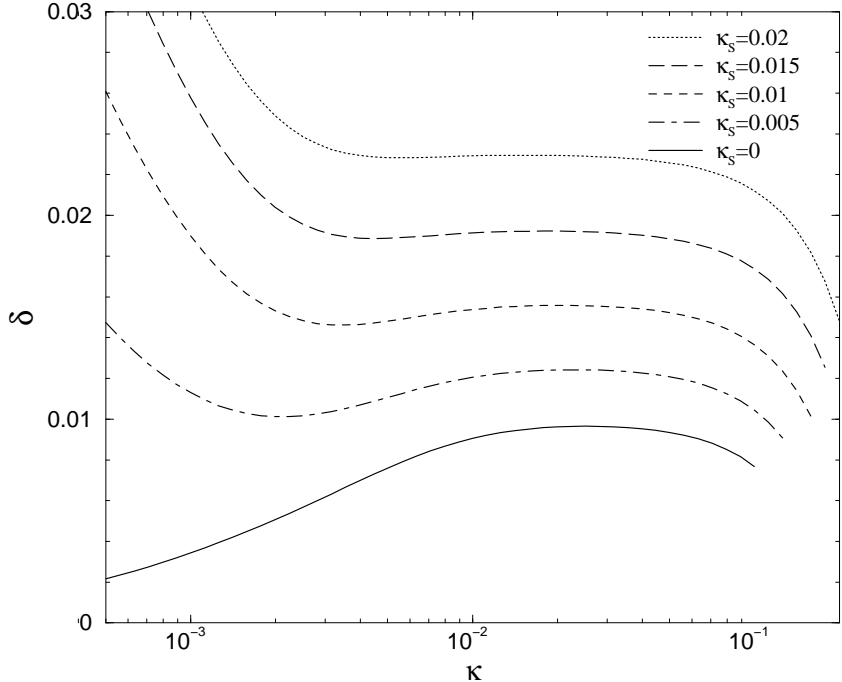}}\hfil
\caption{\footnotesize{Non-Minimal K{\"a}hler potential: predicted
  value of  the 1-loop
  contribution to the spectral index, $\delta$,  depending on the
  value of the coupling $\kappa$, for different values of
  $\kappa_S$; from top to bottom $\kappa_S=0.02,\,0.015,\,0.01,\,0.005,\,
  0$. (${\cal N}=1$).} }
\label{plot2}
\end{figure}

The spectral index is given by:
\be
n_s \simeq 1 -2\kappa_S- 2\delta \,.
\ee
From the previous analysis with minimal K{\"a}hler potential, we could
think naively that the one-loop contribution $\delta \leq 0.01$, and
then we would need for example $\kappa_S \geq 0.01$ if we want the
spectral index around or
below $n_s \approx 0.96$.
However, as previously noted, the non-minimal K\"ahler contribution
will decrease $V^\prime$ which, from Eq. (\ref{spectrum}), tends to
increase the amplitude of the curvature perturbation.
Thus, in order to keep the WMAP  normalization, the scale of inflation
$M$ (i.e. $V$) has to decrease accordingly, see Fig. (\ref{plotM}).  Also, a
decrease in $V^\prime$ means a  smaller value of the field at 50
e-folds. Therefore, for a given value of the coupling $\kappa$, both
the scale of inflation $M$ and the
value of the inflaton field $S_{Re}$ decrease, but in such a way that
their ratio $x_e=S_R/(\sqrt{2} M)$ remains practically
constant\footnote{We have checked this numerically.}. Comparing the
prediction for $\delta$  (Eq. (\ref{etah})) in
the minimal ($\kappa_S=0$) and  non-minimal case ($\kappa_S\neq 0$),
the latter gets enhanced with respect to the minimal case due to the
reduction in $M$.  Therefore,
when taking into account the effects of the non-minimal K{\"a}hler
potential we have also that the 1-loop contribution can be well above
the previous upper limit of $0.01$. Note that there is no
regime where the 1-loop contribution could be neglected with respect
to the non-minimal K{\"a}hler one, as far as the approximation $S_R <
m_\mathrm{P}$ is fulfilled.  This can clearly be seen in
Fig. (\ref{plot2}) where we show  the 1-loop contribution
to the spectral index, $\delta$, for different values of $\kappa_S$.
The general trend is that the 1-loop effective potential contribution
always remains non-negligible, and besides $\delta > \kappa_S$.

\begin{figure}[th]
\hfil\scalebox{0.6} {\includegraphics{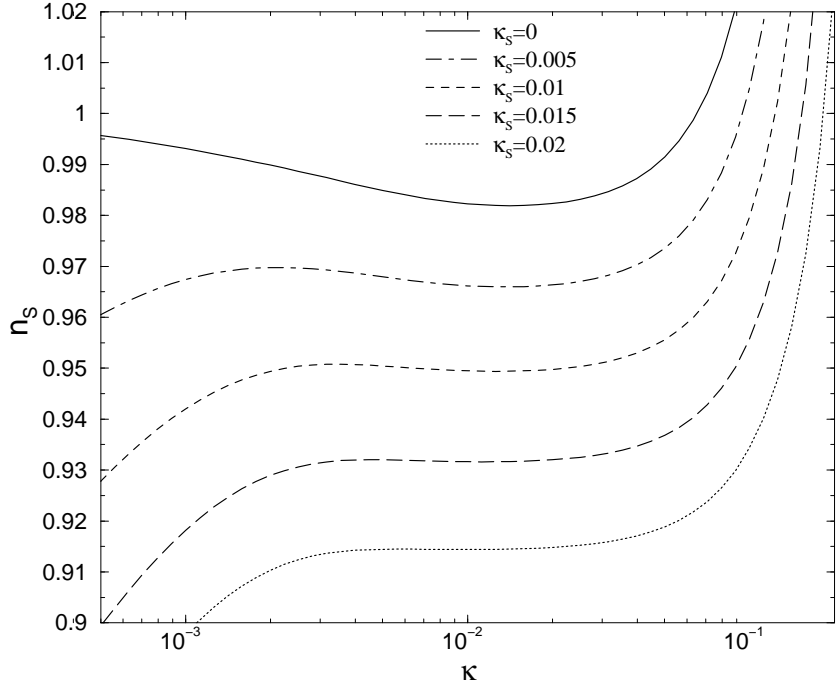}}\hfil
\caption{\footnotesize {Non-Minimal K{\"a}hler potential: predicted
  value of the spectral
  index $n_s$ depending on the
  value of the coupling $\kappa$, for different values of
  $\kappa_S$. We have taken ${\cal N}=1$.} }
\label{plot3}
\end{figure}

In Fig. (\ref{plot3})  we plot the prediction
for $n_s$ as function of $\kappa$ for different values of
$\kappa_S$. We  can see that
even for small values of $\kappa$, already for $\kappa_S
\simeq 5 \times 10^{-3}$ we obtain a spectral index smaller than what we
would have expected only from the non-minimal K{\"a}hler contribution,
due to the increase in $\delta$. As the value
of $\kappa_S$ increases,  the effect gets larger and the spectrum more
and more red-tilted. However, for a given value of $\kappa_S$, the
prediction for the spectral index is practically independent of
the value of $\kappa$, for values of the coupling in the range [0.001,
0.05].

On the other hand, as we increase the coupling and $\kappa
\approx 0.1$, the field value at 50 e-folds also increases and
approaches the Planck scale, as in the minimal case. The
quartic term in the potential then takes over  and gives rise to
a blue-tilted spectrum, just as  with the minimal
K{\"a}hler potential. At which value of $\kappa$ this effect dominates
depends on the value of the quartic coefficient $\gamma_S$, which
in turn may depend now also on the next
parameter in the expansion of the non-minimal K\"ahler potential,
i.e. $\kappa_{SS}$. Nevertheless, for values of $\kappa_{SS} < 1/3$,
this parameter has no effect on the spectral index.

\section{Reheating and Baryon Asymmetry}

To proceed further, an inflationary model should specify the transition
to radiation domination, and also explain the origin of the observed baryon
asymmetry. For hybrid inflation models this has been extensively studied
(see \cite{Senoguz0512} for a review and additional
references). For the non-minimal
models under discussion, let us consider two well motivated examples.
For the first one we identify $G$ with the local $U(1)_{\mathrm{B-L}}$
symmetry, and  introduce three MSSM singlet right-handed neutrinos $N_i$
(i=1,2,3). These acquire masses via the non-renormalizable couplings
\be
 W_2= \frac{1}{m_{\mathrm{P}}} \SuperField{N} \SuperField{N}
 \SuperField{\bar \phi} \SuperField{\bar \phi} \,, \label{WNR}
\ee
where, for simplicity, we will ignore family indices. The couplings
in Eqs (\ref{eq:W1}) and (\ref{WNR}) ensure that the inflaton fields
$\phi$, $\bar \phi$ and  $S$ decay into right handed neutrinos and
sneutrinos. The reheat  temperature is roughly given by
\cite{Senoguz0412, Senoguz0512}
\be
                    T_{R}\simeq (10^{-1} - 10^{-2}) M_N \,,\label{TR}
\ee
where $M_N$ denotes the mass of the heaviest singlet neutrino which
satisfies $2 M_N \leq m_{\mathrm{inf}}= \sqrt{2} \kappa M$. Here
$m_{\mathrm{inf}}$  denotes the inflaton mass in the global minimum
after inflation.
The gravitino constraint usually requires that $T_R \leq 10^6- 10^9$ GeV,
and so non-thermal leptogenesis \cite{nonthermal} is the most
plausible scenario in these
models for generating the observed baryon asymmetry in the universe.

Our second example is based on the flipped $SU(5)$ model. The reason one
avoids GUTs such as $SU(5)$ and $SO(10)$ has largely to do with the primordial
monopoles which appear at the end of inflation and create a serious
cosmological problem. The well known doublet-triplet problem in $SU(5)$ is also
nicely resolved in flipped $SU(5)$. Hybrid inflation in flipped
$SU(5)$ was recently discussed and shown to yield a spectral index
$n_s=0.99 \pm 0.01$ \cite{Kyae0510}. By using a non-minimal K\"ahler
problem we can  obtain
$n_s$ close to 0.95, in much better agreement with the three year WMAP
results. Note that baryogenesis via leptogenesis is also automatic in
flipped $SU(5)$.

In summary, we have argued that a relatively modest extension of minimal
supersymmetric hybrid inflation preserves many of its successful features
and also yields a scalar spectral index which appears to be more 
consistent with the most recent data.

\section*{Acknowledgements}
Two of us (MBG and SFK) would like to thank the CERN TH division
for its hospitality during the writing of this paper. We are grateful to
Nefer Senoguz for his insightful comments and suggestions, and to
Toni Riotto for helpful discussions. Q. S. is supported
in part by DOE under contract number DE-FG02-91ER40626.

%\bibliography{inflation}
%\bibliographystyle{TitleAndArxiv}

\providecommand{\bysame}{\leavevmode\hbox to3em{\hrulefill}\thinspace}

\end{document}